\documentclass[conference]{IEEEtran}
\IEEEoverridecommandlockouts
\usepackage{cite}
\usepackage{amsthm,amsmath,amssymb,amsfonts}
\usepackage{algorithmic}
\usepackage{graphicx}
\usepackage{textcomp}
\usepackage{xcolor}
\usepackage{float}
\usepackage{subfigure}
\usepackage{algorithm}
\usepackage{algorithmic}
\usepackage{bbold}
\usepackage{array}
\usepackage{booktabs}
\usepackage{mathrsfs}
\usepackage[normalem]{ulem}

\usepackage{todonotes}
\usepackage{url}

\def\BibTeX{{\rm B\kern-.05em{\sc i\kern-.025em b}\kern-.08em
    T\kern-.1667em\lower.7ex\hbox{E}\kern-.125emX}}
    
\begin{document}

\title{Realizing XR Applications Using 5G-Based 3D Holographic Communication and Mobile Edge Computing\\
{}
}

\author{
    \IEEEauthorblockN{Dun Yuan\IEEEauthorrefmark{1}, Ekram Hossain\IEEEauthorrefmark{1}, Di Wu\IEEEauthorrefmark{2},  Xue Liu\IEEEauthorrefmark{2},  Gregory Dudek\IEEEauthorrefmark{2}}
    \IEEEauthorblockA{Samsung AI Center Montreal, Canada
    \\\IEEEauthorrefmark{1}\{dun.yuan, e.hossain\}@partner.samsung.com \IEEEauthorrefmark{2}\{di.wu1, steve.liu, greg.dudek\}@samsung.com }
    }


\maketitle

\begin{abstract}
3D holographic communication has the potential to revolutionize the way people interact with each other in virtual spaces, offering immersive and realistic experiences. However, demands for high data rates, extremely low latency, and high computations to enable this technology pose a significant challenge. To address this challenge, we propose a novel job scheduling algorithm that leverages Mobile Edge Computing (MEC) servers in order to minimize the total latency in 3D holographic communication. One of the motivations for this work is to prevent the uncanny valley effect, which can occur when the latency hinders the seamless and real-time rendering of holographic content, leading to a less convincing and less engaging user experience. 
Our proposed algorithm dynamically allocates computation tasks to MEC servers, considering the network conditions, computational capabilities of the servers, and the requirements of the 3D holographic communication application. We conduct extensive experiments to evaluate the performance of our algorithm in terms of latency reduction, and the results demonstrate that our approach significantly outperforms other baseline methods. Furthermore, we present a practical scenario involving Augmented Reality (AR), which not only illustrates the applicability of our algorithm but also highlights the importance of minimizing latency in achieving high-quality holographic views.
By efficiently distributing the computation workload among MEC servers and reducing the overall latency, our proposed algorithm enhances the user experience in 3D holographic communications and paves the way for the widespread adoption of this technology in various applications, such as telemedicine, remote collaboration, and entertainment.
\end{abstract}

\begin{IEEEkeywords}
    Extended reality (XR) applications, holographic communication, mobile edge computing, job scheduling
\end{IEEEkeywords}

\section{Introduction}

Holographic communication is an advanced form of extended reality (XR) technology that involves the real-time capturing, encoding, transporting, and rendering of 3D representations of people, objects, or environments \cite{xu20113d}. In this process, the 3D object is captured using multiple cameras or depth-sensing devices, which generate a volumetric representation of the object. This volumetric data is then encoded using efficient compression techniques to reduce the required bandwidth for transmission over networks such as 5G. Once the data is transmitted, it is decoded and rendered on the user's display device, creating a lifelike, immersive 3D holographic experience. 3D holographic communication allows users to interact with the virtual content as if it were physically present, fostering a more natural and engaging communication \cite{clemm2020toward, akyildiz2022holographic}. The seamless integration of holographic communication into XR applications such as VR, AR, and MR dramatically enhances the user experience by providing high realism, depth perception, and interactivity.

\subsection{Challenges}
The high data rates and computational demands associated with 3D holographic communication impose significant challenges on the underlying communication networks, particularly in terms of latency \cite{petkova2022challenges}. Excessive delays can degrade the quality of holographic content, leading to poor user experience and the uncanny valley effect, a phenomenon where a user perceives the holographic representations as eerie or unsettling due to slight imperfections or inconsistencies \cite{shoydin2022uncanny}. Addressing latency issues is, therefore, crucial for ensuring seamless and real-time holographic experiences.

One promising solution to minimize latency is through the efficient use of Mobile Edge Computing (MEC) servers \cite{mao2017survey, siriwardhana2021survey}. MEC is a paradigm that brings cloud computing resources closer to the end-users by offloading computation tasks from User Equipments (UEs) to nearby edge servers \cite{maray2022computation, chen2018computation}. This not only reduces the communication latency but also allows for more efficient utilization of the available computational resources, leading to a reduction in the overall latency \cite{guo2017energy}. Nevertheless, to fully harness the benefits of MEC in 3D holographic communication, specialized algorithms and techniques for job scheduling and resource allocation are required \cite{wang2022task}.

\begin{figure}[htbp]
\centering
\includegraphics[scale=0.35]{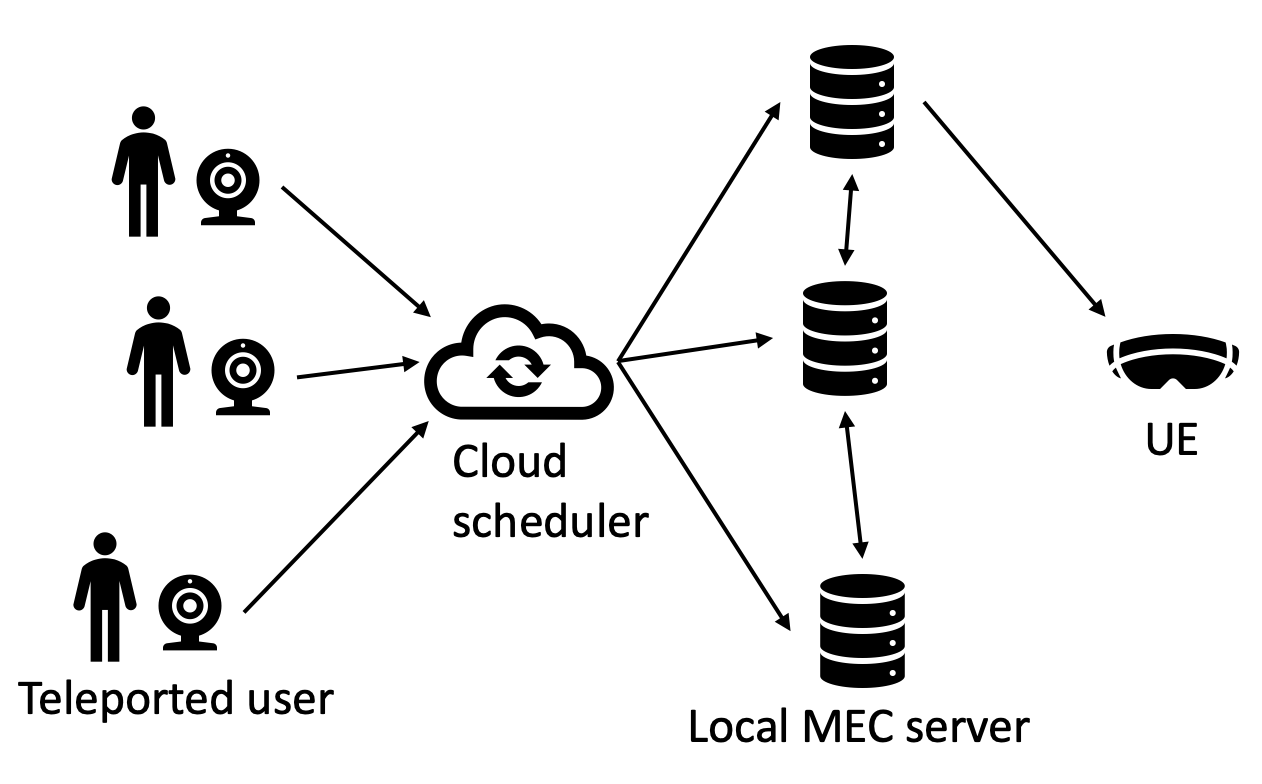}
\caption{Illustration of edge computing-assisted 3D holographic communication.}
\label{Illustration}
\end{figure}

As illustrated in Fig. \ref{Illustration}, the 5G holographic communication process involves several key components, including teleported users, a cloud scheduler, local MEC servers, and UEs. The teleported users generate holographic communication tasks, which are then offloaded to the cloud scheduler. The scheduler, using our proposed algorithm, allocates tasks to the local MEC servers for processing. The processed data is then transmitted back to the UEs for rendering and display.

\subsection{Literature Review}

The development of holographic communication has seen significant progress in recent years, attracting attention from academia and industry. The critical aspects of holographic communication include the compression of digital holographic data \cite{dufaux2015compression}, data transmission requirements \cite{xu20113d}, and potential challenges and solutions in achieving truly immersive holographic communication \cite{clemm2020toward, akyildiz2022holographic}. A comprehensive review of holographic-type communication and its challenges was presented in \cite{petkova2022challenges}. Researchers have looked into 3D holographic transmission in a variety of ways. For instance, the tile-based transmission was explored in \cite{hooft2019tile,xiao2017optile,guo2018optimal}, allowing increased performance and efficiency during the communication of holographic data in wireless networks. Uncanny valley effect is a psychological phenomena with an impact on the final performance of 3D holographic communication that has been researched by several works in different aspects \cite{shoydin2022uncanny, moore2012bayesian}.

MEC plays a crucial role in enabling low-latency and resource-efficient holographic communication. Several surveys \cite{mao2017survey, siriwardhana2021survey} have discussed the communication perspective of MEC, its architectures, and its applications, including mobile augmented reality. Adaptive streaming of holographic content has been studied in \cite{amirpour2020towards}, while efficient rendering techniques for point-sampled geometry have been proposed in \cite{botsch2002efficient}.

IoT-cloud service optimization in next-generation smart environments was discussed in \cite{barcelo2016iot}. The survey of IoT cloud platforms was presented in \cite{ray2016survey}. Joint optimization of computational cost and devices' energy for task offloading in multi-tier edge-clouds was explored in \cite{el2019joint}. Other studies have investigated joint optimization for task offloading in edge computing from various perspectives, including evolutionary game approaches \cite{dong2019joint}, energy-efficient resource allocation \cite{you2016energy}, and multi-user mobile edge computing \cite{guo2017energy}. Task offloading with multi-tier computing resources in next-generation wireless networks was investigated in \cite{wang2022task}. In terms of next-generation wireless networks, 5G-Advanced and the evolution towards 6G have been discussed in \cite{chen20235g}. Studies on centralized and decentralized channel estimation in FDD multi-user massive MIMO systems were presented in \cite{rajoriya2022centralized}. Computation offloading in mobile cloud computing and mobile edge computing was surveyed in \cite{maray2022computation}, while computation peer offloading for energy-constrained MEC in small-cell networks was explored in \cite{chen2018computation}.

5G networks can enable many new applications in different areas. Cellular-connected wireless VR that supported by 5G and beyond with evaluation by QoS has been introduced in \cite{hu2020cellular}. Reference \cite{chen2018virtual} also studied a scenario consisting of VR with data and other information being transmitted through wireless networks. Moreover, fully-decentralized fairness-aware federated MEC small-cell peer-offloading for enterprise management networks was presented in \cite{chi2022fully}, and latency optimization for resource allocation in mobile-edge computation offloading was explored in \cite{ren2018latency}. Remote production for live holographic teleportation applications in 5G networks was discussed in \cite{qian2022remote}, while the "Uncanny Valley" effect in holographic image transmission was investigated in \cite{shoydin2022uncanny}.

\subsection{Contribution}

In this paper, we propose a novel algorithm that focuses on job scheduling by cloud servers to local MEC servers of UEs, with the primary objective of minimizing the total latency in 3D holographic communication. Our approach dynamically allocates computation tasks, taking into account the network conditions, computational capabilities of the MEC servers, and the requirements of the 3D holographic communication application. By efficiently distributing the computation workload among MEC servers and leveraging their proximity to the UEs, our algorithm significantly reduces the overall latency, thereby enhancing the user experience and paving the way for the widespread adoption of 3D holographic communication.

To validate the performance of our proposed algorithm, we conduct a comprehensive set of experiments on a simulated 5G network, including teleported users, the cloud scheduler, local MEC servers, and UEs. The simulation parameters are adopted based on real-world data, ensuring a realistic evaluation of our approach. The simulation results demonstrate that our algorithm outperforms several baseline methods, including local computation by teleported users, join the shortest queue, and always split the task evenly to all MEC servers in terms of latency reduction. These baseline methods are briefly introduced to provide context for comparing our proposed algorithm.

The proposed job scheduling algorithm, which minimizes the total latency in 3D holographic communication,  contributes to the ongoing efforts to develop efficient and scalable solutions for this emerging technology. The improved user experience resulting from our algorithm is expected to foster further adoption of 3D holographic communication across various applications and industries, ultimately transforming the way people interact in virtual environments.

The remainder of this paper is organized as follows: Section II presents the problem formulation on 3D holographic communication. Section III details our proposed algorithm with analysis. Section IV presents the experimental results, and Section V concludes the paper with discussions on future work.

\section{System Model and Problem formulation}\label{method}

\begin{table}[h]
\centering
\caption{Notations}
\begin{tabular}{|c|p{2.5in}|}
\hline
\textbf{Symbol} & \textbf{Description} \\ \hline
$M$ & Set of MEC servers \\ \hline
$N$ & Set of teleported users \\ \hline
$b_{m,n}$ & Bandwidth of link $(m,n)$ \\ \hline
$b_{m_1,m_2}$ & Bandwidth of link $(m_1,m_2)$ \\ \hline
$l_n$ & Latency for teleported user $n$ on UE \\ \hline
$K$ & Set of data classes \\ \hline
$C$ & Set of computing operations \\ \hline
$s_k$ & Size of a data in class $k$ \\ \hline
$t$ & Number of splits by cloud scheduler \\ \hline
$p_{k,c,m}$ & MEC $m$'s computing capacity of processing a class-$k$ data with operation $c$ \\ \hline
$A_{k,c}[t]$ & Number of computing tasks to process class-$k$ data with operation $c$ \\ \hline
\end{tabular}
\end{table}

This section presents our proposed model for the 3D holographic communication system in detail. We will introduce an overview of the pipeline for the whole process, from task and data generation to finally data transmission to UE after rendering. We also explain our proposed job scheduling algorithm for 3D holographic communication with computation on MEC servers. Our approach aims to minimize the total latency by efficiently allocating computation tasks to the MEC servers, considering the network conditions, computational capabilities of the servers, and the requirements of the 3D holographic communication application.

\begin{figure}[htbp]
\centering
\includegraphics[scale=0.35]{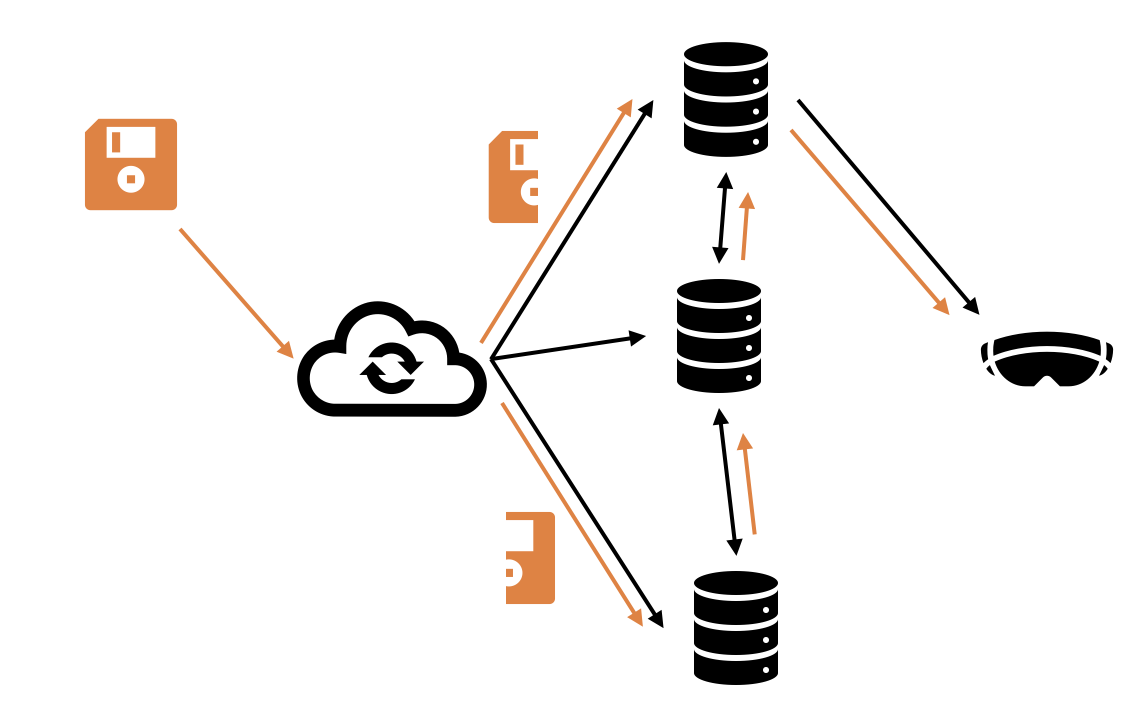}
\caption{Data splitting and processing pipeline of cloud scheduler and MEC servers.}
\label{Detailed}
\end{figure}

Fig. \ref{Detailed} provides a detailed view of the data splitting and processing pipeline, which involves the cloud scheduler and MEC servers. The pipeline consists of the following steps:

\begin{enumerate}
\item\textbf{Task Generation:} Teleported users generate holographic communication data related to the 3D holographic content. The class of data is denoted by $k$ which $k\in K$ and corresponded size of data is $s_k$.

\item\textbf{Task Offloading:} The generated data are offloaded to the cloud scheduler, which is responsible for allocating the tasks to the MEC servers based on our proposed algorithm. The bandwidth of this process is $b_{m,n}$.

\item\textbf{Data Splitting and Allocation:} The cloud scheduler splits the tasks into $t$ smaller subtasks and allocates them to the MEC servers $M$, taking into account the network conditions $b_{m_1,m_2}$, computational capabilities $p_{k,c,m}$ of the MEC servers, and the desired latency $l_n$ of the 3D holographic communication application.

\item\textbf{Task Processing:} The MEC servers process the assigned subtasks, leveraging their computational resources $p_{k,c,m}$ to perform the necessary calculations and rendering $A_{k,c}[t]$.

\item\textbf{Data Integration and Transmission:} The processed $t$ numbers of data from the MEC servers is integrated and synchronized by one of the servers, ensuring that the complete holographic content is transmitted to the UEs.
\end{enumerate}

In terms of the detailed transmitted data through the wireless network, this paper uses point clouds of teleported users as an example. The point cloud of a person includes coordinates of every point that represents the person's shape. The number of points in a point cloud depends on the setup and sensor device. Since we only need to transmit the points of the person rather than the whole scene, the number of points could be significantly reduced. Reference \cite{qian2022remote} provides exact information about the requirements: ultra HD resolution point cloud of one person requires 3756000 points, therefore streaming at 30 FPS would need 2.9 Gbps, which could be supported only by 5G and beyond wireless networks.

Given the notations presented in Table 1, the objective is to minimize the total latency for the single UE and there are multiple teleported users corresponding to this UE. The total latency, $l_n$, for each teleported user $n$ is a combination of communication and computation latencies, which can be represented as:
\begin{equation}
l_n = \sum_{m \in M} \frac{s_k}{b_{m,n}} + \sum_{k \in K} \sum_{c \in C} \frac{A_{k,c}[t]}{p_{k,c,m}}.
\end{equation}

The system requirements for this problem include the following constraints. For each MEC server $m$, the computing capacity $p_{k,c,m}$ must be sufficient to process the class-$k$ data with operation $c$.
The bandwidth of the link between the MEC servers, $b_{m_1,m_2}$, should be sufficient to support the data transfer between them.
The bandwidth of the link between the MEC servers and the teleported users, $b_{m,n}$, should be sufficient to support the data transfer between them.

The problem formulation shows a system that could transfer the data from teleported user side to the cloud scheduler, then to MEC servers. Next, data and corresponding tasks will be split and integrated across servers for an optimization process to improve the performance in terms of minimizing total computation and communication of UE.

\section{Proposed Task Scheduling Algorithm}

The proposed algorithm consists of two linear programs (LPs), LP1 and LP2. The former focuses on finding the minimized maximum latency among users, and at the same time, the latter aims to minimize the number of splits with respect to the maximum latency obtained from LP1. Combining these two LPs allows for a fair allocation of tasks to MEC servers and ensures a balanced Quality of Service (QoS) for all teleported users.

\textbf{Linear Program 1 (LP1):} Find the minimized maximum latency among users and limit the maximum total latency among all teleported users' data to $l_{max}$. The formulation is as follows:
\begin{align*}
\text{minimize } & l_{max} \\
\text{subject to } & l_n \leq l_{max},  \forall n \in N \\
& \sum_{m \in M} \frac{s_k}{b_{m,n}} + \sum_{k \in K} \sum_{c \in C} \frac{A_{k,c}[t]}{p_{k,c,m}} \leq l_n,  \forall n \in N \\
& l_n \geq 0,  \forall n \in N
\end{align*}

\textbf{Linear Program 2 (LP2):} Minimize the number of splits regarding $l_{max}$, which means they cannot exceed $l_{max}$ but allow the same or lower latency. The objective of minimizing the number of splits is to decrease synchronization and integration overhead, also increasing the robustness of the overall system. The formulation is as follows:
\begin{align*}
\text{minimize } & \sum_{n \in N} t_n \\
\text{subject to } & l_n \leq l_{max},  \forall n \in N \\
& \sum_{m \in M} \frac{s_k}{b_{m,n}} + \sum_{k \in K} \sum_{c \in C} \frac{A_{k,c}[t]}{p_{k,c,m}} \leq l_n,  \forall n \in N \\
& l_n \geq 0, t_n \geq 0,  \forall n \in N 
\end{align*}

Running LP1 and LP2 would result in a fair allocation of tasks to MEC servers and fair QoS regarding the performance on all teleported users. The reason for using two linear programs instead of only optimizing the final total latency is to consider the fairness and balancing of load for all teleported users. If only one teleported user exists, the algorithm would be the same as only running LP1, and LP2 would not affect the result. If multiple teleported users exist, the LP2 part of the algorithm achieves a balance of total latency among all teleported users while keeping the maximum latency below a target value. In this way, all teleported users share the exact value of total latency, which lowers the overhead on task splitting and data synchronization, therefore, benefits the performance on an even larger scale.
\section{Experimental Results}\label{exp}

\subsection{Experimental Setup}

Our simulation setup is designed to evaluate the performance of the proposed job scheduling algorithm. It has been tested in a 3D holographic communication simulator that emulates a 5G New Radio wireless network. The network contains components based on the problem formulation. The main components of the simulation include teleported users, a cloud scheduler, local MEC servers, and UE. The teleported users generate holographic communication tasks, which are offloaded to the cloud scheduler. Using the proposed algorithm, the cloud scheduler allocates tasks to the local MEC servers for processing, considering the servers' network conditions and computational capabilities. We adopt realistic parameters for the 5G network based on real-world data, ensuring a reliable evaluation of our approach. The key parameters include bandwidth, processing capacity, latency, task sizes, and processing time. In the experiments, we use three MEC servers as local MEC servers for offloading tasks. These three servers will cooperate to fulfill the performance requirements of the 3D holographic communication for UE according to the tasks scheduled by the cloud scheduler. 

\subsection{Evaluation Metrics}

\textbf{Total Latency}: The primary evaluation metric used in our experiments is the total latency, which includes both communication and computation latency. This latency has an intuitive inverse relationship with the Quality of Service (QoS) of the system.


However, we also acknowledge the importance of considering the uncanny valley effect \cite{mori2012uncanny}, which introduces a more sophisticated relationship between likability, which serves as the final performance score, and the resemblance to the teleported user.

\textbf{Likability Performance Score}:  We adopt a curve from previous research work \cite{mori2012uncanny} that investigates this phenomenon. By combining this curve with the total latency results, we create a more reasonable final goal metric for evaluating the performance of our proposed algorithm and the baseline methods.

Despite the potential for higher latency to result in better likability performance scores  when using the non-linear curve adopted from uncanny valley research, our main focus remains on minimizing the total latency to ensure a seamless holographic experience. Consequently, in the results section below, we will present both the total latency and the performance score based on this metric, showcasing the effectiveness of the proposed algorithm and the baseline methods in minimizing latency and enhancing the user experience.

\subsection{Baselines}
We introduce three baseline algorithms used for comparison with our proposed job scheduling approach. These algorithms represent common strategies for handling computation tasks in MEC-based 3D holographic communication systems.

\textbf{Local Computation:} Teleported users perform all the computation and rendering tasks locally on their PCs or servers. Once the tasks are completed, the results are sent to the MEC servers solely for synchronization and integration. While this method reduces the load on the MEC servers, it may lead to increased latency due to the limited computational resources of the users' local devices and the need to send results to the MEC servers.

\textbf{Join Shortest Queue:} The cloud scheduler server does not split any incoming data or tasks. Instead, it sends tasks to the MEC server with the fewest computation jobs in its queue. This approach aims to minimize the computation overhead associated with synchronization and integration. However, it may result in suboptimal resource allocation and increased latency if some MEC servers become overloaded while others need more utilized.

\textbf{Always Split Evenly:} The cloud scheduler server consistently splits incoming tasks and data evenly across all MEC servers. This ensures that each server is allocated an equal number of tasks, thereby promoting fair resource distribution. However, this approach does not take into consideration the varying computational capabilities of the MEC servers or the specific requirements of the 3D holographic communication application. Consequently, it may not always lead to optimal latency reduction.

These three baseline algorithms provide a reference for evaluating the effectiveness of the proposed job scheduling algorithm.

\subsection{Simulation Results}

\begin{table}[h]
\centering
\caption{Average total latency of different methods}
\begin{tabular}{|c|c|}
\hline
\textbf{Method} & \textbf{Average total latency (ms)} \\ \hline
Proposed (per user)& 364 $\pm$ 47 \\ \hline
Local Computation & 932 $\pm$ 53 \\ \hline
Join Shortest Queue (per user) & 646 $\pm$ 61 \\ \hline
Always Split Evenly (per user)&  730 $\pm$ 38\\ \hline
\end{tabular}
\end{table}

\begin{figure}[htbp]
\centering
\includegraphics[scale=0.5]{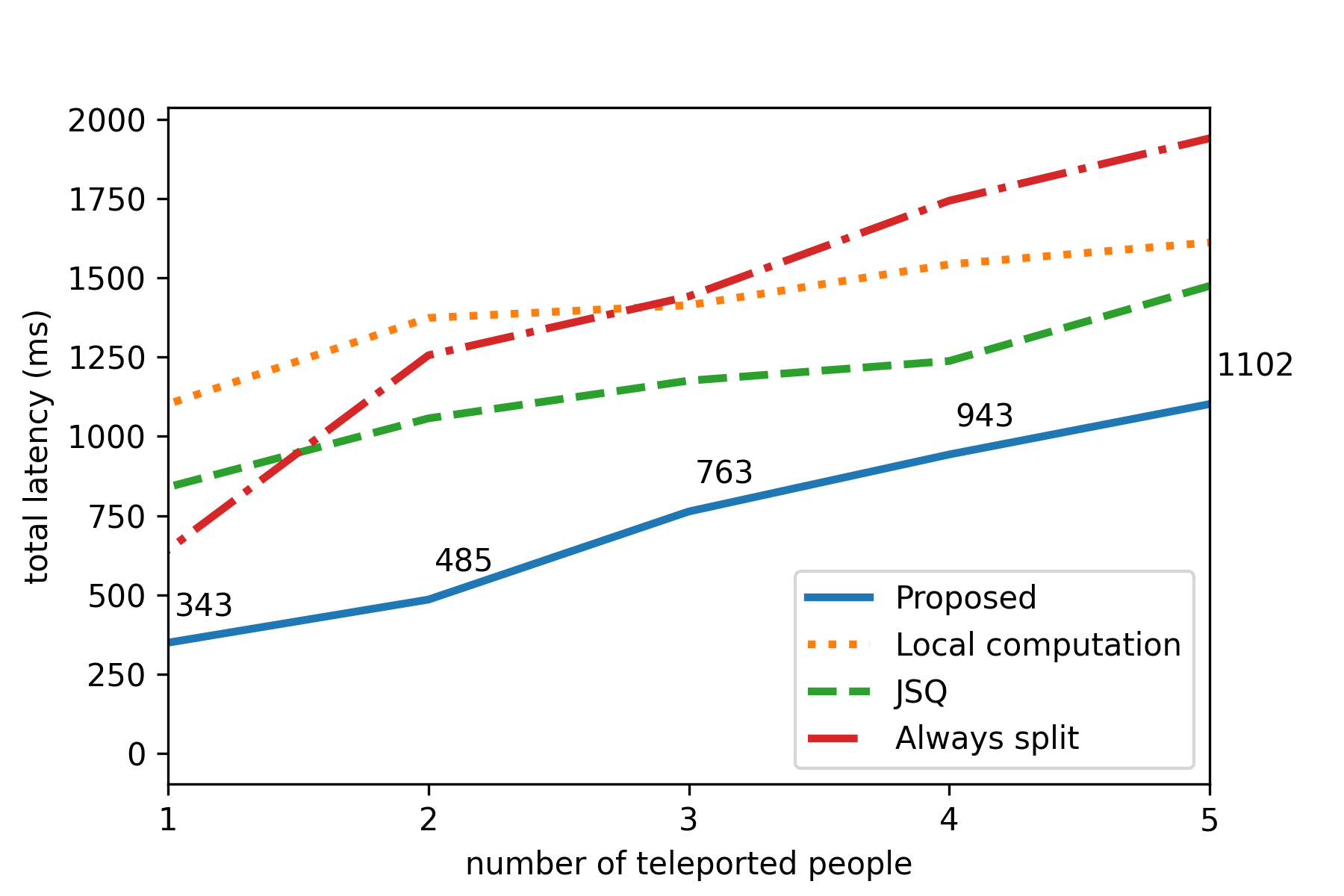}
\caption{Results from different algorithms in terms of total latency (ms).}
\label{Results1}
\end{figure}

The results in Fig. \ref{Results1} demonstrate that the proposed job scheduling algorithm outperforms all the baseline methods in terms of latency reduction. Specifically, we observe a significant decrease in the total latency when using our algorithm, as compared to the baseline methods. This indicates that our approach efficiently allocates tasks to the MEC servers and optimizes their utilization, leading to an overall improvement in the performance of the 3D holographic communication system.

In addition to the latency results, we also evaluate the likability performance score as a metric to better understand the user experience in the context of the uncanny valley effect. The performance score provides a comprehensive measure of the holographic experience, capturing the impact of latency on the system's likability and realism.

\begin{figure}[htbp]
\centering
\includegraphics[scale=0.5]{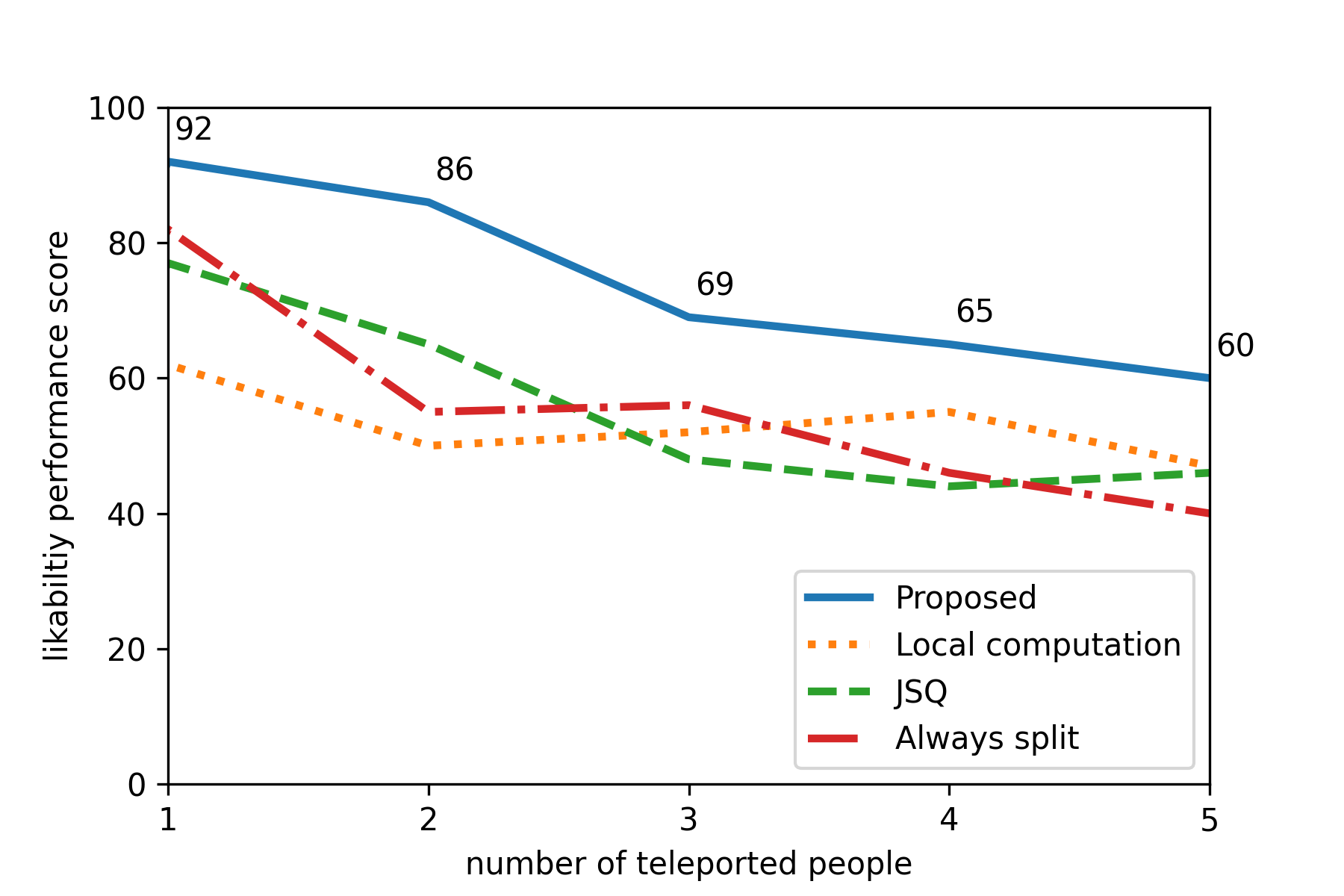}
\caption{Results from different algorithms in terms of likability performance score.}
\label{Results2}
\end{figure}

As depicted in Fig. \ref{Results2}, the proposed algorithm outperforms the baseline methods in terms of the likability performance score, which combines the total latency results with the impact of the uncanny valley effect. This demonstrates that our approach effectively mitigates the uncanny valley effect by minimizing latency, leading to a more seamless and realistic holographic experience for users. Specifically, our algorithm achieved a 36\% reduction in total latency and a 28\% increase in the likability performance score compared to the best performing baseline method. The superior performance of our algorithm in this regard highlights its practical value and applicability in real-world 3D holographic communication applications.

\subsection{Demonstration}

\begin{figure}[htbp]
\centering
\includegraphics[width=0.5\columnwidth]{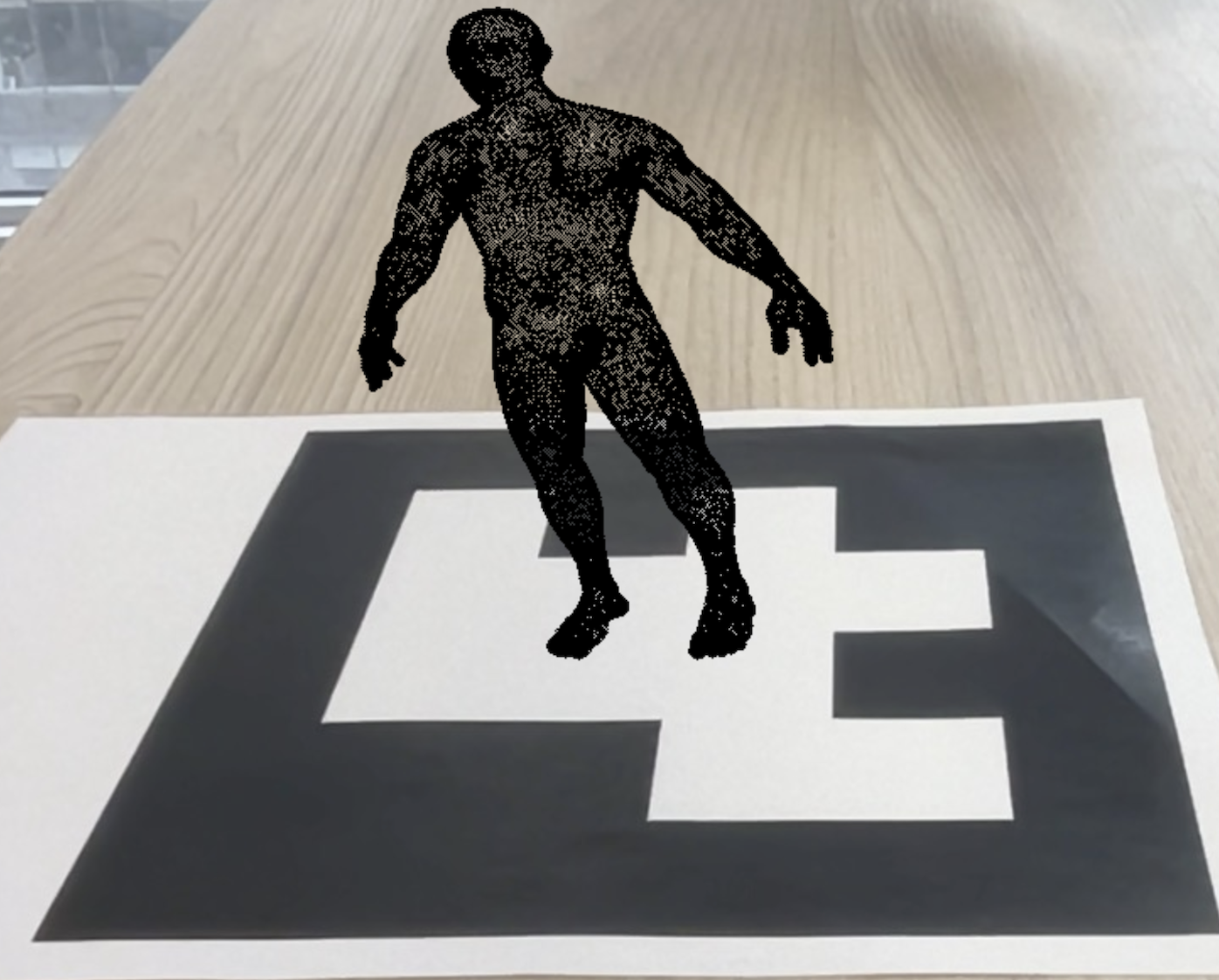}
\caption{A human model hologram constructed by a point cloud. Simulated in 3D holographic communication.}
\label{result3}
\end{figure}

In addition to the quantitative results, we present an illustrative demonstration of 3D holographic communication shown in Fig. \ref{result3}. This demonstration showcases the practical implications of our algorithm and emphasizes the importance of minimizing latency for achieving high-quality holographic views. The demonstration is conducted on a simulated 5G network. It features a real-time augmented reality scene where users can interact with holographic objects and observe the achievability of the 3D holographic teleportation approach. The demonstration results highlight the effectiveness of our proposed algorithm in enhancing the user experience by reducing latency and preventing the uncanny valley effect in practical scenarios.

Our experiments provide strong evidence for the performance of our proposed job scheduling algorithm in minimizing the total latency for 3D holographic communication, and in terms of likability performance score.

\section{Conclusion}\label{conclusion}
We have presented a novel job scheduling algorithm for cloud servers to local MEC servers in the context of 3D holographic communication. The primary objective of it is minimizing the total latency to improve the performance with uncanny valley taken into consideration. Our algorithm significantly reduces overall latency by dynamically allocating computation tasks, considering network conditions, MEC server computational capabilities, and the 3D holographic communication application requirements. This improvement results in an enhanced user experience and promotes the widespread adoption of 3D holographic communication across various applications and industries. Our comprehensive experiments on a 5G network simulator have demonstrated that the proposed algorithm outperforms several baseline methods regarding total latency reduction and likeability performance score, combining the uncanny valley curve and total latency. Furthermore, we have presented an illustrative demo showcasing the practical applications of our algorithm and the importance of minimizing latency for an optimal holographic experience.

In our future work, we plan to extend our algorithm to consider multiple UEs and investigate fairness between them. It will enable our approach to address more complex, real-world scenarios and further improve the performance of 3D holographic communication systems. Additionally, we aim to further refine our evaluation methodology to better capture the relationship between latency and user experience in 3D holographic communication.




{
\bibliographystyle{IEEEtran}
\bibliography{ref.bib}
}
\end{document}